\documentclass[11pt,twoside]{article}

\usepackage{asp2014}

\aspSuppressVolSlug
\resetcounters

\begin{document}

\title{The Heliophysics Coverage Registry: An integrated metadata system for coordinated, multi-mission solar observatories}

\author{Neal~Hurlburt,$^1$  Ryan~Timmons,$^1$ and Ralph Seguin$^1$}
\affil{$^1$Lockheed Martin Advanced Technology Center, Palo Alto, CA, USA; \email{hurlburt@lmsal.com}}

\paperauthor{Neal~Hurlburt}{hurlburt@lmsal.com}{}{LMATC }{LMSAL}{Palo Alto}{CA}{94304}{USA}
\paperauthor{Ryan~Timmons}{timmons@lmsal.com}{}{LMATC }{LMSAL }{Palo Alto}{CA}{94304}{USA}
\paperauthor{Ralph~Seguin}{seguin@lmsal.com}{}{LMATC }{LMSAL }{Palo Alto}{CA}{94304}{USA}

\begin{abstract}
Modern studies of the Sun involve coordinated observations collected from a collage of instruments on the ground and in orbit. Each instrument has its own constraints, such as field of view, duty cycle, and scheduling and commanding windows, that must both be coordinated during operations and be discoverable for analyses of the resulting data. Details on the observed solar features, i.e. sunspots or filaments, and solar events, i.e. flares or coronal mass ejections, are also incorporated to help guide data discovery and data analysis pipelines. The Heliophysics Coverage Registry (HCR) provides a standards-based system for collecting and presenting observations collected by distributed, ground and space based solar observatories which form an integrated Heliophysics system. The HCR currently supports all instruments on the Interface Region Imaging Spectrograph (IRIS) and Hinode missions as well as associated ground-based observatories. Here we present an overview of the HCR along with details on how it provides scientists with tools to make flexible searches on observation metadata in coordination with searches of solar features and events.
\end{abstract}

\section{Introduction}
Like other areas of astronomy, Solar Physics has come to rely more and more on coordinated observations involving multiple sites, spacecraft and instruments. Each instrument and site has unique capabilities (e.g. field of view, waveband) and behaviors (e.g. maximum cadence, orbital constraints) that needs to taken into account while planners are attempting to capture solar intermittent features and explosive events, and which needs to be discoverable once the data is captured in the archive. 
\section{The Hinode Observation System: 2006}
The basic concepts for the HCR were first implemented in the Hinode Observation System (HOS) which became operational in 2007. The system is based on a modified VOEvent scheme. Predicted events are created during observation planning on the Chief Observer Workstation (COWS), where the basic elements of the observation are captured. Data flowing down from the Hinode Spacecraft is cataloged and assembled into Observation events. These are correlated by the Event Processor, along with commentary from members of the science team. Finally the results are distributed via various routes.

\subsection{The Heliophysics Coverage Registry:2010}
The HOS was generalized and incorporated into the Heliophysics Event Knowledgebase (HEK) as the Heliophysics Coverage Registry (HCR). Data collected by an instrument inherits properties both of the instrument and how it is operated. This includes observatory properties, such as where it is located (ground or space) and what observational constraints exist (day/night cycle, seeing conditions, etc.); details on how the instrument is used and properties of the instrument itself, including instrument effects. 

\section{Integration with the Heliophysics Events Knowledgebase}
\articlefigure[width=.5\textwidth]{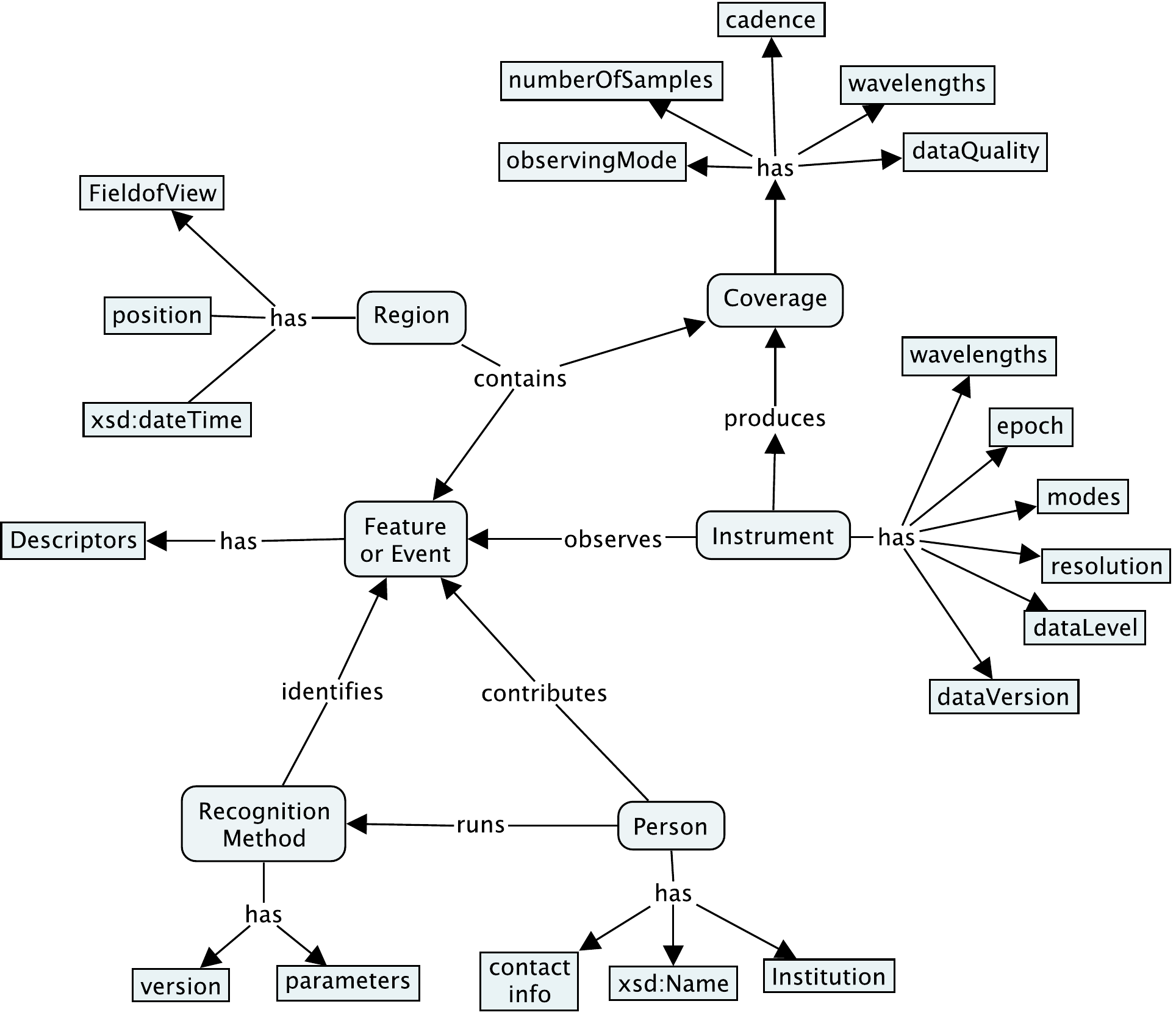}{fig1}{The HCR is integrated into the HEK to provide a faceted search capability.}

Data flowing from the Solar Dynamics Observatory's \citep{2012SoPh..275....3P} AIA and HMI instruments are analyzed by a dozen feature detection algorithms which report their results into the Heliophysics Events Registry (HER), along with events found from other data sources. These solar events are then available to guide researchers to the appropriate research datasets. To aid that search, the HCR is tightly integrated with the HER as seen in \ref{fig1}~and together they form the Heliophysics Events Knowledgebase \citep{2012SoPh..275...67H} (HEK, not to be confused with the Hunt for Exo-moons by Kepler). The integrated systems enables faceted search capability, where queries can find data sets associated with particular solar events and features or find features contained within a specific dataset.

\section{Upgrade for Interface Region Imaging Spectrograph (IRIS)}
The HCR was revamped in 2013 in part to support close integration with the IRIS \citep{2014SoPh..289.2733D} data system. The API was expanded and generalized and a new search tool was built to take advantages of the changes. The new heksearch tool supports faceted searches keyed on solar events, or particular instruments. These searches are automatically cross-referenced with the other facets with an estimate of coverage overlap. 

Each selected observation presents a series of links to the corresponding dataset. This supports links to external sites where data can then be further filtered or, in the case of IRIS, to direct links of to prepackaged, compressed datasets.

\section{The HCR Today}
The HCR contains almost two decades of observations from TRACE and IRIS Small Explorer missions, all three Hinode mission instruments and extracted data cubes from the AIA instrument on the Solar Dynamics Observatory. Over 165,000 descriptions of solar observations are currently available. The vast majority of these have been viewed at least once, with over 2.5 million views in  total.

\begin{table}[!ht]
\caption{Tables in \LaTeXe}
\smallskip
\begin{center}
{\small
\begin{tabular}{llc}  
\tableline
\noalign{\smallskip}
Observations between 1999-01-01 to 2017-10-01 & Views since 2007-01-01 \\
\noalign{\smallskip}
\tableline
\noalign{\smallskip}
Total number of  events& 165,579  \\
Total number of viewed events &161,086 \\
Total number of views & 2,513,954 \\
\noalign{\smallskip}
\tableline\
\end{tabular}
}
\end{center}
\end{table}

\section{Usage Analysis}
\articlefiguretwo{P10-68_f2.pdf}{P10-68_f3.pdf}{fig2}{Analysis of the access patterns of the HCR.  \emph{Left:} Observation events sorted in rank order shows power law relations.  \emph{Right:} Histograms of the events show the distribution shift to higher number of viewings with more recent missions.}

The top figure shows the number of viewings ordered by rank for each instrument. Prior to the HCR, TRACE data access displayed a 1/ra dependence on ranking with $a=0.5$, suggesting that use of the data was concentrated towards a few observations. 
The other instruments show a behavior fit well be a more general description with view, $V \propto (\max(r)+1-r)^b/r^a$ where $r$ is the rank and $a$ and $b$are constants. The best fit for IRIS is where $a=b=0.3$. The Hinode/SOT and SDO/AIA instruments show more complexity with a high count for the first few observations dropping to a lower curve for the rest. The high values all come from early in the two missions and before we revised the search interface. The lower value of a suggests that these data are being more throughly analyzed.

Histograms of the number of observations with a given number of viewings display another distinction between the pre- and post-IRIS upgrade behaviors. Prior to 2013, all observations from each instrument were most frequently viewed between one and 10 times. IRIS shows a clear shift toward more frequent views, with the median closer to 100. This increase can also be seen as an extended tail to the other distributions.

\section{Conclusion}
The Heliophysics Coverage Registry provides a rich search capability for solar physics data discovery and fusion. Its effectiveness can be seen in the flattening of the access patterns seen in the observation rankings and by the order of magnitude increase in the mean number of views for a given observation. These changes appear to be enabled by providing an advanced, faceted search interface. 

The HCR and its concepts is extensible and we intend to continue supporting new missions as the opportunity arises. Events from external missions are already supported and new ones can be added by creating HCR-compliant VOEvents.

\acknowledgements This work has been supported by NASA funds through  Hinode/SOT and XRT, SDO/AIA and IRIS program funds. We thank Ted Tarbell, Bart de Pontieu, Mark Cheung, Sam Freeland, Greg Slater and Phil Shirts for their help.

\bibliography{P21-48}  

\end{document}